\documentclass[12pt]{article}

\usepackage{graphicx}%
\usepackage{multirow}%
\usepackage{amsmath,amssymb,amsfonts}%
\usepackage{amsthm}%
\usepackage{amssymb}
\usepackage{mathrsfs}%
\usepackage[title]{appendix}%
\usepackage{xcolor}%
\usepackage{textcomp}%
\usepackage{manyfoot}%
\usepackage{booktabs}%
\usepackage{algorithm}%
\usepackage{algorithmicx}%
\usepackage{algpseudocode}%
\usepackage{listings}%
\usepackage{hyperref}
\usepackage{tabularray}
\usepackage{comment}

\usepackage{subcaption} 
\newif\ifsupplement
\supplementfalse   

\newenvironment{sciabstract}{%
\begin{quote} \bf}
{\end{quote}}

\definecolor{forestgreen}{rgb}{0.13, 0.55, 0.13}

    \title{Entropy and diffusion characterize mutation accumulation and biological information loss}

\author
{Stephan Baehr,$^{1\ast\dagger}$ Hans Baehr,$^{2,3,4\ast\dagger}$\\
\\
\normalsize{$^{1}$Biodesign Center for Mechanisms of Evolution, School of Life Sciences,} \\ \normalsize{Arizona State University, USA}\\
\normalsize{$^{2}$Department of Physics and Astronomy,} \\ \normalsize{University of Georgia, Athens, GA 30602, USA}\\
\normalsize{$^{3}$Center for Simulational Physics, University of Georgia, Athens, GA 30602, USA}\\
    \normalsize{$^{4}$Max Planck Institute for Astronomy, Königstuhl 17, D-69117 Heidelberg, Germany}\\
\\
\normalsize{$^\ast$E-mail: sbaehr@asu.edu, baehr@mpia.de} 
\\
\normalsize{$\dagger$ These authors contributed equally to this work.}
}

\date{}

\begin{document}


\baselineskip24pt

\ifsupplement\else
\maketitle 

\section*{Abstract:}
\begin{sciabstract}

Aging is a universal consequence of life, yet researchers have identified no universal theme. This manuscript considers aging from the perspective of entropy, wherein things fall apart. We first examine biological information change as a mutational distance, analogous to physical distance. In this model, informational change over time is modeled by an advection-diffusion equation, in this case a normal distribution with a time component. The solution of the advection-diffusion equation provides a means of measuring the entropy of diverse biological systems. The binomial distribution is also sufficient to demonstrate that entropy increases as mutations, epimutations, or other biological errors accumulate. As modeled, entropy scales with lifespans across the tree of life. This perspective provides potential mechanistic insights and testable hypotheses as to how evolution has attained enhanced longevity: entropy management. We find entropy is an inclusive rather than exclusive aging theory.
 
\end{sciabstract}







\section*{Introduction}
\label{sec:intro}

The biology of aging can be described by the phrase, ``things fall apart". Researchers have noted that though there is rhyme and similarity to aging among individuals, each case is unique and unprogrammed \cite{finch_chance_2000, kirkwood_programmednon-programmed_2011, alon_systems_2024}. Leonard Hayflick \cite{hayflick_entropy_2007} has argued for decades that aging is entropy, an increase in molecular disorder over time. The concept thus far has only enjoyed modest popularity, perhaps because it does immediately offer a direct means of measurement, treatment, or a specific molecular mechanism. References to entropy in aging research are often vague and allusory, offering few testable hypotheses. In principle, the phenomena that researchers have been measuring all along, and which are known to modify lifespan, should also fit naturally within the variables that describe entropy; this has yet to be emphatically shown. The purpose of this manuscript is to make a bridge between the world that the biology of aging knows and measures, and the physical understanding of entropy.   

Superficially, the signature of entropy whittling at organism genomes across time may be recognized as the accumulation of deleterious mutations. As deleterious mutations accumulate, information disperses. Mutation accumulation is notably a familiar concept to aging biology, being among the oldest concepts in the field in one form or another.\cite{medawar_unsolved_1952, szilard_nature_1959, orgel_maintenance_1963}. The idea of accumulating errors leading to critical breakdowns of biological systems is referred to as a ``error catastrophe" or ``mutation catastrophe".  A modern version of mutation catastrophe is supported by some evidence; DNA mutations do accumulate over time in cells over a lifespan.\cite{milholland_mutation_2017,vijg_dna_2021, cagan_somatic_2022}. However, we note that mutation accumulation need not only refer to DNA, if the definition is broadened: what is epigenetic information loss, if it is not the accumulation of (epi)mutations, and what of everything else going awry in cells in aging? While acknowledging the importance of DNA mutations in the aging process \cite{lopez-otin_hallmarks_2023}, for example in the emergence of cancer \cite{tomasetti_variation_2015}, the field has been emphasizing that epimutations likely also have a proximal role to play in both aging and cancer. Compounding evidence \cite{rando_aging_2012} and recent experimentation \cite{yang_loss_2023} have highlighted the need for a model that includes the importance of more than just DNA mutation. Ideally, a theory should be flexible to account for aging of all sorts, even for organisms that age over the course of days, such as in \emph{E. coli}.

We propose a model where the accumulation of mutations over time between at least two points can be considered a ``mutational distance". We fit the concept of mutational distance, analogous to ``hamming" distance\cite{adami_evolution_2024}, to physics definitions of distance via an advection-diffusion equation for the Brownian motion, and use the result to model the change in entropy of a system over time. From a starting point of highly similar cells within a population, the cells accumulate mutational distance over time. The model fits to DNA mutation accumulation experiments. We then model entropy as a primary factor in the determination of longevity, though the role of DNA mutation and all other systems of entropic gain. We fit the model to organisms of varying lifespan and demonstrate the model's flexibility, which predicts that an entropic failure threshold causes biological mortality, via age-related phenotypes. We also examine a simple binomial entropy conversion for diverse biological systems, and its application to age-related molecular change. These simplified models suggest that aging may be entropy; and that entropy also increases within germline lineages as well, in the relative absence of selection.

\section*{Results}
\subsection*{Gaussian systems}
The inspiration for this work begins by recognizing that the phenotypic outcomes of biological aging are shared with those of evolutionary biology's mutation accumulation (MA) experiments. In MA experiments, an increasing burden of random mutations results in phenotypic degradation\cite{kibota_estimate_1996, ajie_behavioral_2005}, because the average DNA mutation is deleterious \cite{eyre-walker_distribution_2007}. The sum of mutations per line is counted to provide an estimate for mutation rate and mutation burden per line. This mutation accumulation, or mutational distance, over evolutionary time and in the presence of selection leads to the differences that define individuals, populations, and species. At baseline, however, unchecked mutation accumulation leads to mutational meltdown, phenotypic degradation, and lineage extinction. 

Within a population of cells, in an MA or within an aging soma, Figure \ref{fig:advecdiffusion} examines the behavior of mutations within a population of cells over time. When interpreting mutational distance as physical distance, this perspective allows the import of physics equations that model evolution over time, with particular consideration of Brownian motion, and random walks of molecules. The model considers mutation accumulation as a one-dimensional distance and how a population of molecules will be distributed as a function of time, following a normal distribution. The equation for a normally distributed variable $x$ is:
\begin{equation} \label{eq:normaldistribution}
f(x,\mu,\sigma) = \frac{1}{\sigma \sqrt{2\pi}} e^{ -\frac{(x - \mu)^2}{2\sigma^2} }
\end{equation}
where $\mu$ is the distribution mean and $\sigma^2$ is the variance. 

An advection-diffusion equation is often used to model the diffusive spread of a quantity, for example a drop of food coloring in a stream, or a rubber duck race down a river, \emph{Entenrennen} in German, considering the one-dimensional distance from a starting point. Figure \ref{fig:advecdiffusion}A demonstrates the one-dimensional distribution of rubber duckies floating down a river with some current: the mean distance increases as a function of time, with some spread in the distribution, which follows a normal distribution. This solution to the advection-diffusion equation is simply a normal distribution with a time component $t$, diffusion coefficient $D$, current or flow rate $D\lambda$, and drag coefficient $\lambda$ (more information in Supplemental Information: Models).
\begin{equation} \label{eq:advecdiffsolution}
F(x,t) = \frac{1}{\sqrt{4\pi Dt}} e^{-(x - D \lambda t)^2/4Dt}.
\end{equation}
We note that as an MA experiment proceeds, the spread of the distribution widens as a function of the mean, which is conventionally characterized by a Poisson distribution. The widening of the distribution is well appreciated \cite{lee_rate_2012} and because the normal distribution is a good approximation of the Poisson distribution for a large enough Poisson mean, the normal distribution suffices. We test the advection-diffusion model upon real MA data arising from both WT and hypermutator \emph{E. coli} \cite{lee_rate_2012, baehr_consideration_2025} in Figure \ref{fig:advecdiffusion}B. Figure \ref{fig:advecdiffusion}B demonstrates that a reasonable goodness of fit is approximated by the solution to the advection-diffusion equation. The $D\lambda$ variable contains the fold-difference between the wild-type and hypermutator strains; about 110-fold, and may be considered analogous to the current or flow rate. The variable $\lambda$ is a fitting parameter analogous to a drag coefficient, which helps fit the observed variance to the mean distance from a starting point of zero mutations.

\begin{figure*}[!t]
\centering
\includegraphics[width=0.88\textwidth]{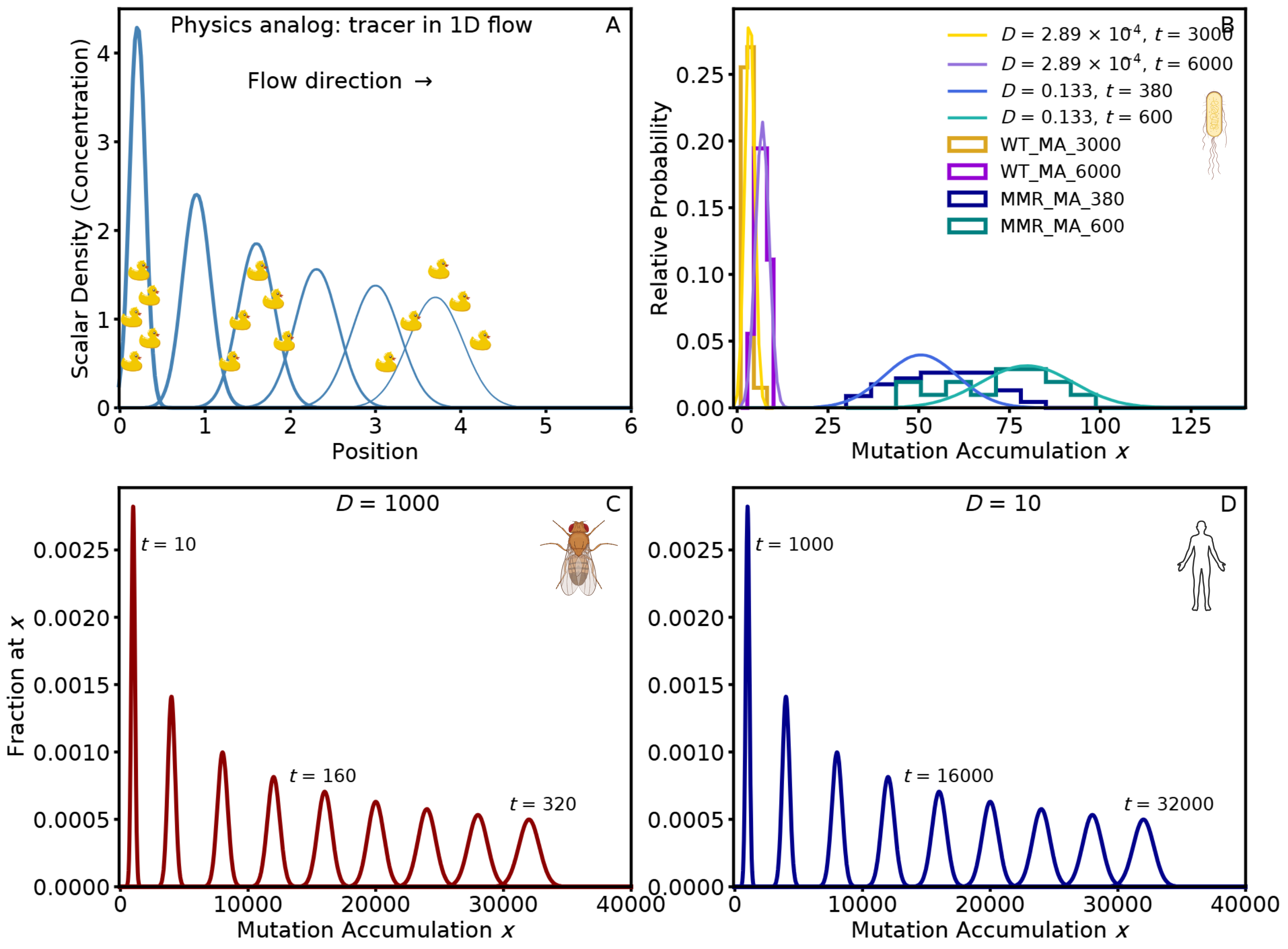}
\caption{{\bf The advection-diffusion solution as a function of time and mutation accumulation.} Panel A: A simple example of how a 1D diffusion model can model a passive tracer in a fluid, in this case, rubber duckies floating down a river. Panel B: The advection-diffusion equation applied to \emph{E. coli} samples. Panels C and D: The advection-diffusion equation models how diverse organism lifespans may end up with the same final result. If chromatin “drift” or “epimutation” are primary drivers of aging, this equation is sufficiently flexible to model it.}
\label{fig:advecdiffusion}
\end{figure*}

In addition to DNA mutations, the term ``epimutation" is used to describe changes in informational content of an epigenome, and epimutations accumulate. For researchers who are more interested in epimutations than DNA mutations in the context of aging, the advection-diffusion model remains capable of quantifying the increasing disorder gained among a population of cells accumulating epimutations. Using epimutation burden estimates obtained from DNA methylation burden\cite{bertucci-richter_rate_2023} as a coarse example, advection-diffusion model can be used to model epimutations. In Figure \ref{fig:advecdiffusion}C and \ref{fig:advecdiffusion}D, the model demonstrates that the same increase-in-variance outcome can be obtained over orders of magnitude of time-frame, for example from humans to \emph{D. melanogaster}. The advection-diffusion equation is incredibly flexible, and is a suitable model to describe informational distance change across biological informational storage media. We simply use a rough DNA methylation average epimutation distance as a proof-of-principle for a broader epimutation rate argument; it is well known that \emph{D. melanogaster} lacks DNA methylation.\cite{gibert_paramount_2021}

A useful insight of the advection-diffusion equation is that it is Gaussian in nature, and from a Gaussian time series an estimate of the variance in terms of $D$ and $t$ can be achieved; which is independent of $\lambda$. This result, $\sigma^2=2Dt$, further expounded upon in the supplemental methods section, can be applied to an equation for the entropy of a Gaussian distribution,Equation S10. The substitution yields an equation for the entropy of a Gaussian distribution in terms of $D$ and $t$, Equation \eqref{eq:shannongaussian}. 

\begin{equation} \label{eq:shannongaussian}
H = \frac{1}{2} (\ln (4\pi Dt) + 1).
\end{equation}
In principle, any biological process with a normal distribution that is subject to change over time can be converted into units of entropy. As variance increases, so too does the entropy within a gaussian system.

\subsection*{Binomial systems}
Entropy can be calculated in diverse contexts. As described by Manfred Eigen in 1971\cite{eigen_selforganization_1971} and summarized more recently\cite{adami_evolution_2024}, mutation itself absent selection is an event of entropic gain within a system. It is easy enough to state that \textit{information has been lost} when an \emph{E. coli} genome acquires a random mutation, but the statement is quantifiable. Using an information theoretic construction, when a cell acquires a mutation, the mutation could be anywhere in the genome and be of any base-substitution. The more mutations that are acquired, the more combinations are possible which could describe the original ancestor state, when agnostically looking back in time from a derived mutation-accumulated state. For example, taking only a single cell of an aging tissue or from the \emph{E. coli} MA, entropy can be quantified by the amount of information lost from acquiring 1, 2, or 80 random mutations. Under the assumption that each mutation has an average effect and a simplified model assuming only two states, mutated or not, the entropy can be quantified by the binomial:
\begin{equation}
W = \binom{n}{k} = \frac{n!}{k!\,(n-k)!} = C(n,k)
\end{equation}
Where $n$ is the number of mutations within a genome and $k$ describes the genome size. The solution of this equation further resolves into entropy via the unitless entropy equation (absent a Boltzmann constant)
\begin{equation}
S = \ln(W)
\end{equation}
Where S is the entropy of a system in nats, and W is the number of microstates possible from the accumulation of mutations in the \emph{E. coli} genome. In the MA, the entropy becomes 15, 28, and 954 nats, respectively. At this time, is unclear which measure of entropy, population level or within an individual cell, is more appropriate or relevant. It is also unclear as to what degree the entropy of a system needs to be scaled to make the measures equivalent.

DNA nucleotides are not simply binomially state-described as ``mutated" or ``not mutated", but there is a particular intent and usefulness of the binomial treatment: flexibility. Specifically, researchers studying the biology of aging have acquired some skill classifying certain age-related changes as "bad" from a "good" starting state. With regard to proteins, the accumulation of oxidative damage is deleterious, protein misfolding and its accumulation is deleterious, protein mis-localization is deleterious. By considering the most simple state of entropy, the binomial, the hallmarks of aging\cite{lopez-otin_hallmarks_2023} recognizably fall into some measure of entropy. 

\subsection*{Sum-total measures of entropy}
There is a clear space in aging analysis for a parallelization of multiple entropy parameters within a single cell or organism. The entropic gain of DNA mutation, transcript error, gene expression noise, translation error, chromatin structure, and even things like protein misfolding, or oxidative damage to proteins or lipids may in principle be modeled by some form of entropy. We summarize this perspective to incorporate all possible sources of entropy (i.e. DNA mutation, transcript error, translation error, chromatin structure, etc.)
\begin{equation} \label{eq:total_entropy}
H_{\text{total}} = \sum_i H_i .
\end{equation}
where $H_i$ is each potential source of entropy. At this time it is unclear as to how the varying system's entropic gains should be normalized; perhaps by some factor of the relative fitness cost per unit of entropy gained for each system measured. We simply note that the sum of all entropy within a cell or group of cells is relevant to encountering an entropic threshold, which we consider the ``entropy catastrophe".

\begin{figure}[!t]
\centering
\includegraphics[width=0.88\textwidth]{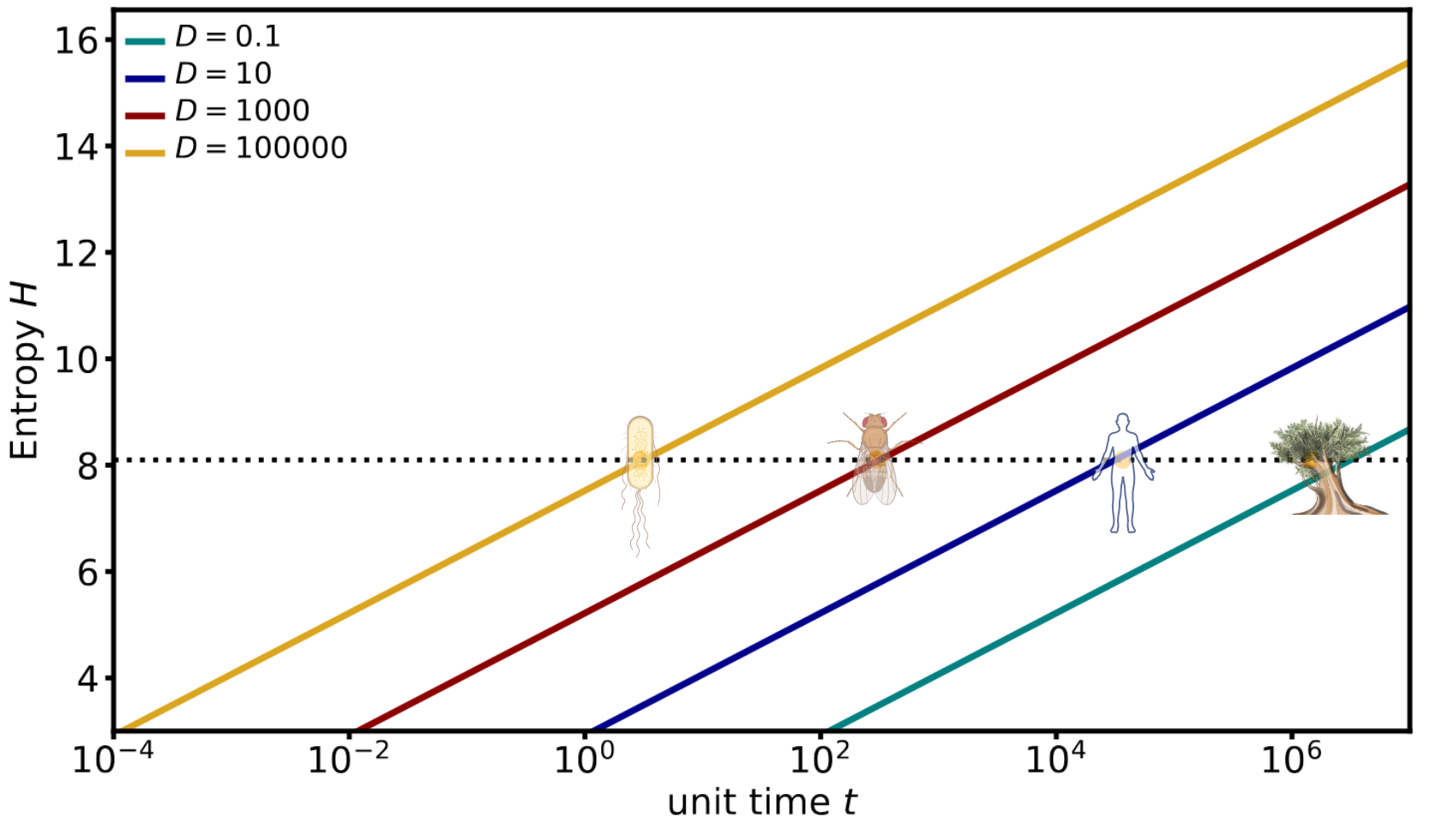}%
\caption{{\bf Entropy and diffusion scale across orders of magnitude of time}. The total entropy of a system will depend strongly on the value of the diffusion $D$. Orange points mark the rough maximum lifespans of a few organisms (from left to right): \emph{E. coli}, \emph{D. melanogaster}, \emph{homo sapiens} and \emph{pinus longaeva}, assuming our unit of time is in days, as listed in Supplementary Table 2.}
\label{fig:entropycomparison}
\end{figure}

Under a threshold model of system failure, such as has been recurrently proposed in biology of aging research \cite{orgel_maintenance_1963, milholland_mutation_2017}, we hypothesize that organisms that reach their mortality sooner have higher must -total rates of entropy gain. One way to vary the rate of entropy gained in a system is by varying diffusion rates, $D$, which equivocate to mutation rates, over time. To this end, we extrapolate from our existing estimates of $D$ (Figure 1C, 1D) and append short-lived \emph{E. coli}\cite{wang_robust_2010} and long-lived Bristlecone pine trees \emph{Pinus longaeva} to demonstrate the effect of rate of variance spread upon estimates of entropy. The entropy equation models a log-linear relationship of increasing entropy with time. We note that the value of $D$ extrapolated for \emph{E. coli} in Figure \ref{fig:entropycomparison} and the one directly calculated from DNA mutation accumulation data are highly divergent from one another; this may reflect the idea that \emph{E coli's} DNA mutation rate is sufficiently low, relative to its lifespan, that DNA mutation is incredibly unlikely to contribute to the replicative senescence experienced by the bacterium.

Diffusion is also a parameter of the binomial distribution. The entropy of the normal distribution is a function of its variance ($\sigma^2$), as seen in Equation S10. The entropy $H$ of the binomial distribution, which is approximated by a gaussian distribution, may be calculated as found in Equation S9, in which diffusion may be put in terms of probability. Therefore, by multiple routes, the general trend of Figure 2 remains consistent.

\section*{Discussion}
\label{sec:discussion}

The perspective of this manuscript examines a bridge between the biology of aging, physics, and evolutionary biology. This perspective began as a study of the phenotypic similarities emergent in mutation accumulation experiments and of the biology of aging, or specifically, an inescapable degradation of biological phenotypes in both contexts.

By recognizing that mutational distance is equivalent to physical distance, equations of physics may be applied to biological data. The advection-diffusion equation models the dispersion of molecules from a starting state of identical position, to a probability distribution of distance over time. This dispersion can be readily appreciated in the acquisition of DNA mutations over the course of human development and aging: all cells start out with an identical genotype, but over time a clock-like distance from the starting state to an aged state emerges over time. The model predicts an increased dispersion of mutational distance in chromatin or in DNA, though we also focus upon chromatin and its information storage role. Importantly, we find that the advection-diffusion equation can be modeled to fit mutation accumulation data from \emph{E. coli} reasonably well, and can resolve the expected deviation in mutation rate between wild-type and hypermutator strains. The key insight of the advection-diffusion equation is its provision of a description of the variance of the Gaussian distribution in an age-related context.

Mathematicians and physicists have long recognized the importance of the Gaussian distribution to the study of entropy. The equation for entropy in a Gaussian system is straightforward via Equation \eqref{eq:shannongaussian}); the parameter of importance is the variable $D$ in Equation \eqref{eq:advecdiffsolution}. The model compares which values of $D$ might give rise to known lifespan estimates across the tree of life. The model indicates that the organisms are hitting a similar entropy threshold for varying levels of $D$. By combining the perspectives of Figure \ref{fig:advecdiffusion} and Figure \ref{fig:entropycomparison}, a threshold model with a mean entropy acquisition rate over time, the model satisfyingly predicts even organisms with identical starting genotypes would reflect a Gaussian distribution of survivorship over time, focused about a mean. This is the result of aging experimentation on genetically identical organisms.\cite{finch_chance_2000}

The above perspective provides some avenues of application. The model proposed here, which we call the ``entropy catastrophe hypothesis" for the biology of aging, provides testable hypotheses. Specifically, the hypothesis predicts that variance should increase over time, at least in the biological systems that are causal to aging. The accumulation of DNA mutations and epimutations are part of the hypothesis, but remain only two facets among the broad story of aging. The hypothesis predicts that evolutionary innovations that reduce entropy, such as increased replication fidelity, the induction of recycling programs, rewriting/restarting programs, and inducing purifying selection upon deleterious subsets of molecules, are responsible for enhanced longevity among organisms. The model proposes that interventions that increase the rate of aging, such as stress, temperature, or conditions like Hutchinson-Gilford progeria syndrome, ultimately act by increasing entropy in the system at biologically relevant levels.

There is undoubtedly a differential contribution to aging phenotypes and mortality, from differing molecular biological systems. Even within biological information systems, it likely true that the `fitness impact' of epimutations in somatic cells is far less per mutation than that of DNA. Judging by relative mutation rates of the molecules as a proxy for relative importance, it may be that epimutations, 100 to 1000-fold more prevalent than DNA mutations, need to be weighted correspondingly such that each epimutation is 'worth' an inverse proportion to their prevalence; if not so extreme, it is certainly in that direction. To clarify, the reason translation or transcription errors are orders of magnitude more prevalent\cite{lynch_evolutionary_2023} than DNA is almost assuredly because their individual impact is all that much less important than a single DNA mutation. The idea that chromatin information in mammals is the weak link is attractive to the field, but remains to be proven. For the present moment, we note that diffusion and entropy likely characterize the distributions of diverse age-related molecular phenomena. Their ultimate impacts, perhaps species and context specific, remain to be resolved.

\section*{Methods}
\section*{Figures}

Figure 1 has been generated in panel A by simple application of the advection-diffusion equation. Panel B experimental datasets are derived from the MA experiments reported in 2012\cite{lee_rate_2012} and 2025\cite{baehr_consideration_2025}, whose data have been transformed/rearranged into Supplementary Table 2. Data transformation and re-arrangement requires binning mutation counts per line into sub-groups.

Panel C and D of Figure 1 are estimates of epimutation rates derived from experimental results from 2023\cite{bertucci-richter_rate_2023}; approximately 1\% of DNA methylation sites are becoming discordant, or mutated, over one lifespan in several mammals. For a similar amount in humans, $2.8\times10^7$ CpG sites results in ~$2.8\times10^5$ epimutations in a lifespan. If around 1-10\% of those epimutations might be functional, the functional epimutational burden of DNA methylation alone may be on the order of 2,800 to 28,000. In contrast, the somatic DNA mutational burden of humans is on the order of 3,000-5,000, of which 1\%, or 30-50 mutations are functional. As an aside, the fitness effect-size and distribution of fitness effects for the epigenome are as yet unknown to our knowledge; but evolutionary theory predicts that they will be far less impactful, given their evolutionary impermanence. However, we may make an estimate of these based on their relative prevalence, and guess their average cellular fitness effect may be between 1/100th and 1/1000th that of the average DNA mutation. We acknowledge that the numbers offered are estimates, but offer the model as a general hypothesis to be tested. For simplicity of the model, this result has been translated into a relative number of human epimutations per aged cell; we suggest that the same principle and perhaps order of magnitude should extend also to chromatin marks, generally. Extending the epimutation hypothesis to \emph{D. melanogaster} necessarily requires considering chromatin marks beyond DNA methylation, as flies lack DNA methylation. 

Supplementary Table 1 provides the order-of-magnitude estimates of parameters that gave rise to Figure 2; few \emph{Drosophila} have ever measured to have a maximum lifespan of 100 days, but it is within a factor of 2 of reported values. 
 
\section*{Data Availability Statement} The code used for the analysis of the data is available https://github.com/hbaehr/entropy. The data used for the analyses may be found in the Supplementary Information.

\section*{Competing Interests} The authors declare no competing financial or non-financial interests.

\section*{Acknowledgments}

The authors acknowledge their parents, Wolfgang Baehr and Jeanne Frederick. The authors also seek to acknowledge their mentors, Michael Lynch and Hubert Klahr; and also their communities, scientific and otherwise, for their congenial atmosphere and \emph{tolerance}. This work has been funded in part by NIH GMS grant 5R35GM122566-08 and the National Science Foundation, DBI-2119963, 2021-2026, BII: Mechanisms of Cellular Evolution. Figures 1 and 2 were created with the assistance of Biorender. Figure 2 was created with the assistance of Man Lin.

\bibliographystyle{science}
\bibliography{entropyaging2}

\fi

\end{document}